\documentclass[12pt]{article}
\usepackage{color,amssymb,epsfig,verbatim,amsmath}
\newtheorem{theorem}{Theorem}[section]
\usepackage[title]{appendix}

\def\log{\hbox{log}}

\def\boxit#1{\vbox{\hrule\hbox{\vrule\kern6pt
          \vbox{\kern6pt#1\kern6pt}\kern6pt\vrule}\hrule}}
\def\refhg{\hangindent=20pt\hangafter=1}
\def\refmark{\par\vskip 2mm\noindent\refhg}

\def\refhg{\hangindent=20pt\hangafter=1}
\def\refmark{\par\vskip 2mm\noindent\refhg}

\def\bse{\begin{eqnarray*}}
\def\ese{\end{eqnarray*}}
\def\be{\begin{eqnarray}}
\def\ee{\end{eqnarray}}
\def\bq{\begin{equation}}
\def\eq{\end{equation}}
\def\bse{\begin{eqnarray*}}
\def\ese{\end{eqnarray*}}

\begin{document}
\thispagestyle{empty}

\hfill\today \\ \\

\baselineskip=28pt
\begin{center}
{\LARGE{\bf A note of feature screening via rank-based coefficient of correlation}}
\end{center}
\baselineskip=14pt
\vskip 2mm
\begin{center}
Li-Pang Chen\footnote{\baselineskip=10pt Corresponding Author: Departmant of Statistical and Actuarial Sciences, University of Western Ontario,\\
1151 Richmond St., London, ON, N6A 3K7, Canada, lchen723@uwo.ca}

\end{center}
\bigskip

\vspace{8mm}

\begin{center}
{\Large{\bf Abstract }}
\end{center}
\baselineskip=17pt
{
Feature screening is useful and popular to detect informative predictors for ultrahigh-dimensional data before developing proceeding statistical analysis or constructing statistical models. While a large body of feature screening procedures has been developed, most of them are restricted on examining either continuous or discrete responses. Moreover, even though many model-free feature screening methods have been proposed, additional assumptions are imposed in those methods to ensure their theoretical results. To address those difficulties and provide simple implementation, in this paper we extend the rank-based coefficient of correlation proposed by Chatterjee (2020) to develop feature screening procedure. We show that this new screening criterion is able to deal with continuous and discrete responses. Theoretically, sure screening property is established to justify the proposed method. Simulation studies demonstrate that the predictors with nonlinear and oscillatory trajectory are successfully detected regardless of the distribution of the response. 
}

\vspace{8mm}

\par\vfill\noindent

\par\medskip\noindent
\underline{\bf Short title}: XI-SIS

\clearpage\pagebreak\newpage
\pagenumbering{arabic}

\newlength{\gnat}
\setlength{\gnat}{22pt}
\baselineskip=\gnat

\clearpage

\section{Introduction}

%{\color{red}point biserial correlation}: examine correlation of binary and continuous variables.
%\url{https://en.wikipedia.org/wiki/Point-biserial_correlation_coefficient}

%\url{https://medium.com/@outside2SDs/an-overview-of-correlation-measures-between-categorical-and-continuous-variables-4c7f85610365}

%\citet{Chatterjee:2020} \citep{Chatterjee:2020}.

In the era of big data, ultrahigh dimensional data has become available
from many scientific research fields, including biology, genetics, and finance. Even though a large amount of data is able to be collected, few of them are really informative. In statistical analysis or data science, our interests include model construction or prediction. Before doing so, it is crucial to capture relevant information. In ultrahigh dimensional data analysis, {\it feature screening} is one of the important tools to detect informative variables and remove irrelevant ones.

The key idea of feature screening is to measure the correlation of the response ($Y$) and the predictor ($X$) and then select important predictors by choosing large values of correlation. In terms of linear predictors, i.e., the relationship between $Y$ and $X$ is linear, Fan and Lv (2008) proposed to adopt the Pearson correlation to screen out informative predictors, and such an idea was extended to generalized linear models (e.g., Fan et al. 2009; Fan and Song 2010). Li et al. (2012a) applied Kendall $\tau$ to develop the robust rank correlation. However, those methods may be restrictive in the presence of a nonlinear relationship between the predictor. To address this concern, several advanced feature screening methods have been developed. For example, Fan et al. (2011) proposed nonparametric independent screening by using B-spline basis. He et al. (2013) and Wu and Yin (2015) explored nonparametric regression model with heterogeneous errors. Li et al. (2012b) proposed the model-free feature screening method by using the distance correlation, which can be expressed in terms of the Pearson correlation coefficient. Xia and Li (2020) developed the copula-based partial correlation method. Although early methods are claimed to be valid to deal with arbitrarily distributed responses, there was no theoretical justification to support this claim. In addition, continuous responses seem to be only setting in their simulation studies. On the other hand, regarding discrete responses, Mai and Zou (2013) considered binary responses and proposed Kolmogorov filter method; Sheng and Wang (2020) focused on multi-class responses and developed model-free feature screening method to select variables. From the theoretical perspective, however, those methods may require more additional assumptions to ensure sure screening property. Besides, some methods may have complicated computational procedures because of the involvement of nonparametric settings. Therefore, it motivates us to develop a ``unified'' feature screening criterion that is not only able to deal with arbitrarily distributed responses but also enjoys theoretical results as well as the easy implementation of computations.

%Moreover, even though early methods were able to deal with nonlinear predictors, they may be restricted in monotonic functions (see Section~\ref{Num-study} for details) because they directly measure $Y$ and $X$. As a result, when nonlinear functions are oscillatory, it is possible to fail to detect the corresponding predictors.}

In 2020, Chatterjee (2020) proposed a new rank-based correlation coefficient. Different from early approaches that may have complicated computations, this new method provides simple implementation as the classical coefficients like Pearson's correlation or Spearman's correlation. In addition, this new coefficient is not only able to measure the nonlinear relationship between $Y$ and $X$ but also enjoys several theoretical results such as consistency and asymptotic distribution. In the exploration of real data analysis, Chatterjee (2020) also applied this new coefficient to select predictors. While the numerical results look satisfactory,  we are wondering if this new coefficient is suitable in feature screening. Moreover, from the theoretical perspective, we may ask if feature screening based on this new coefficient still enjoys sure screening property just like early work. Based on these motivations, to answer these questions, in this paper we aim to extend this new coefficient to develop feature screening. Specifically, we rigorously verify that this coefficient is able to measure the correlation between binary response and predictors. Furthermore, we establish sure screening property for the proposed feature screening procedure. Finally, numerical studies verify that the proposed method successfully identifies truly important predictors for continuous/binary responses and outperforms other competitive methods.

The remainder is organized as follows. In Section~\ref{Notation}, we introduce some notation, regression models, and crucial problems that we are going to deal with. In Section~\ref{Method}, we briefly review a rank-based correlation coefficient and show that such a type of coefficient is able to measure the correlation of continuous/binary responses and predictors. After that, we extend such a coefficient to develop the feature screening procedure to detect truly important predictors in ultrahigh-dimensional data. Moreover, the sure screening property is also established to justify the proposed method. In Section~\ref{Num-study}, we design simulation studies to conduct the performance of the proposed method as well as compare the proposed method with its competitors. In Section~\ref{rda}, we demonstrate two real data examples that contain continuous and binary responses, respectively. We conclude the article with discussions in Section~\ref{Summary}. Derivation of theoretical results is placed in Appendix~\ref{pf}.

\section{Notation and Models} \label{Notation}

Let $X \triangleq (X_1,\cdots,X_p)^\top$ denote a $p$-dimensional vector of predictors and let $Y$ be a univariate variable. Without loss of generality, we standardize $X_k$ so that $E(X_k)  = 0$ and $\text{var}(X_k)  = 1$ for $k=1,\cdots,p$. For $i=1,\cdots, n$, $(X_i,Y_i)$ has the same distribution of $(X,Y)$. Here we focus on the case of $p \gg n$. When $Y$ is continuous, $Y$ and $X$ are formulated by the following regression model:
\begin{eqnarray} \label{model}
Y = f_1(X_1) + f_2(X_2) + \cdots + f_q(X_q) + \epsilon,    
\end{eqnarray}
where $\epsilon$ denotes the error term, $f_1,\cdots,f_q$ are unknown functions and $q<n$. On the other hand, if $Y$ is the binary variable, a (generalized) logistic regression is commonly used to characterize $Y$ and $X$:
\begin{eqnarray} \label{logit}
\text{logit}\left\{\text{pr}(Y=1)\right\} = f_1(X_1) + f_2(X_2) + \cdots + f_q(X_q).
\end{eqnarray}
Models (\ref{model}) or (\ref{logit}) can be referred to generalized additive models (GAM) (e.g, Wood 2017), and each predictor has nonlinear relationship with the response. If $f_j(\cdot)$ for $j=1,\cdots,q$ is reduced to be a linear parameteriztion, say $X_j \beta_j$, then (\ref{model}) or (\ref{logit}) are reduced to linear or logistic models.

In addition, from (\ref{model}) and (\ref{logit}), we realize that $X_1,\cdots,X_q$ are informative predictors and are dependent on the response $Y$ in nonlinear forms among all $p$ predictors. As a result, before implementing statistical methods to analyze (\ref{model}) or (\ref{logit}), it is crucial to identify those $q$ predictors from the original $p$ predictors.

\section{The Methodology and Main Results} \label{Method}

\subsection{A rank-based coefficient of correlation} \label{discuss-xi}

In this subsection, we overview the new form of correlation coefficient proposed by Chatterjee (2020). For the $k$th predictor $X_k$, the correlation coefficient between $X_k$ and $Y$ is given by
\begin{eqnarray} \label{xi-new-corr}
\xi(X_k,Y) = \frac{\int \text{var}\left[ E\left\{I(Y \geq t) | X_k \right\} \right] d\mu(t)}{\int \text{var} \left\{I(Y \geq t) \right\}  d\mu(t)},
\end{eqnarray}
where $I(\cdot)$ denotes the indicator function and $\mu(\cdot)$ is the law of $Y$. Moreover, as pointed out by Chatterjee (2020), (\ref{xi-new-corr}) is as simple as those for the classical coefficients, such as Pearson’s correlation, that are used to measure the dependence between two continuous variables. Moreover, $\xi(X_k,Y)$ is indeed in an interval $[0,1]$ because $\text{var}\left[ E\left\{I(Y \geq t) | X_k \right\} \right]  \leq  \text{var} \left\{I(Y \geq t) \right\}$ for every $t$. In particular, if $X_k$ and $Y$ are independent, then $\xi(X_k,Y) = 0$ due to that $E\left\{I(Y \geq t) | X_k \right\}$ is a constant; if $Y$ is a measurable function of $X_k$, then $E\left\{I(Y \geq t) | X_k \right\} = I(Y \geq t)$, yielding $\xi(X_k,Y) = 1$. More detailed properties of (\ref{xi-new-corr}) can be found in Chatterjee (2020).

Additionally, we claim that (\ref{xi-new-corr}) is able to measure the correlation of a categorical variable and a continuous variable as well. To see this, we consider $Y \in \{0,1\}$. Then the numerator of (\ref{xi-new-corr}) becomes 
\begin{eqnarray} \label{binary-numerator}
\int \text{var}\left[ E\left\{I(Y \geq t) | X_k \right\} \right] d\mu(t) 
%%%%%
&=& \text{var}\left[ E\left\{I(Y \geq 0) | X_k \right\} \right] + \text{var}\left[ E\left\{I(Y \geq 1) | X_k \right\} \right] \nonumber \\
%%%%%
&=& \text{var} \left\{ \pi(X_k) \right\},
\end{eqnarray}
where the last equality is due to that $E\left\{I(Y \geq 0) | X_k \right\} = 1$ and $E\left\{I(Y =1) | X_k \right\} = \text{pr}(Y = 1 | X_k) \equiv \pi(X_k)$ with $\pi(\cdot)$ being a unknown link function.  On the other hand, the denominator of (\ref{xi-new-corr}) is rewritten as
\begin{eqnarray} \label{binary-denominator}
\int \text{var} \left\{I(Y \geq t) \right\}  d\mu(t)
%%%%%%%
&=& \text{var} \left\{I(Y \geq 0) \right\} + \text{var} \left\{I(Y \geq 1) \right\} \nonumber \\
%%%%%%%
&=& \text{var} (Y) \nonumber \\
%%%%%%%
&=& \text{pr}\left(Y=1\right)  \text{pr}\left(Y=0\right),
\end{eqnarray}
where the second step is due to $\text{var} \left\{I(Y = 1) \right\} = 0$, the third step is due to the definition of the variance.

Combining (\ref{binary-numerator}) and (\ref{binary-denominator}) with $\pi(X_k)$ specified by $X_k$ and incorporation of empirical estimates, we can obtain the empirical estimate of $\xi(X_k,Y)$:
\begin{eqnarray} \label{xi-binary}
\widehat{\xi}_\text{\tiny binary} (X_k,Y) = \left. \frac{\sum \limits_{i=1}^n \left(X_{ik} - \overline{X}_k \right)^2}{n} \right/ \frac{n_1 n_0}{n^2},
\end{eqnarray}
where $n_y$ is the sample size in the class $Y=y$ and $\overline{X}_k = n^{-1} \sum \limits_{i=1}^n X_{ik}$. Moreover, by simple algebra, (\ref{xi-binary}) can be rewritten as
\begin{eqnarray*}
\left( \overline{X}_{1k} - \overline{X}_{0k} \right)^2 r_\text{pb}^{-2},
\end{eqnarray*}
where $r_\text{pb} = \frac{\overline{X}_{1k} - \overline{X}_{0k}}{S_{X_k}} \sqrt{\frac{n p_0 p_1}{n-1}}$ is called \textit{the point-biserial correlation coefficient}, which is used to measure the correlation between a continuous variable $X_k$ and a binary variable $Y$, with $S_{X_k}^2 = (n-1)^{-1} \sum \limits_{i=1}^n \left( X_{ik} - \overline{X}_k \right)^2$, $p_0 = \frac{n_0}{n}$, $p_0 = \frac{n_1}{n}$ and $\overline{X}_{yk}$ is the mean of the predictors in the class $Y=y$. This result indicates that point-biserial correlation coefficient is treated as a special case of (\ref{xi-new-corr}).

\subsection{Feature Screening Method}

Let 
\begin{eqnarray*}
\mathcal{I} = \left\{ k :  X_k\ \text{is dependent on the response} \ Y  \right\}
\end{eqnarray*}
denote the \textit{active set} containing all relevant predictors for the response $Y$ with size $q = \left|\mathcal{I}\right|$ and $q < n$, and $\mathcal{I}^c$ is the complement of $\mathcal{I}$ which contains all irrelevant predictors for the response $Y$. In addition, for $k=1,\cdots,p$ let $\omega_k \triangleq \xi(X_k,Y)$ denote the correlation coefficient between $X_k$ and $Y$. For $i=1,\cdots,n$, denote $Y_{(i)}$ as the rearranged response according to the sort of the $k$th predictors $X_k$, i.e., $(X_{k,(1)}, Y_{(1)}), \cdots, (X_{k,(n)}, Y_{(n)})$ with $X_{k,(1)} \leq X_{k,(2)} \leq \cdots \leq X_{k,(n)}$ and $X_{k,(j)}$ being the $j$th sorted predictor in $X_k$. The corresponding estimator of $\omega_k$ is given by (Chatterjee 2020)
\begin{eqnarray*}
\widehat{\omega}_k = 1 - \frac{n \sum \limits_{i=1}^{n-1} \left| r_{i+1} - r_i \right|}{2 \sum \limits_{i=1}^n \ell_i \left( n - \ell_i \right)},
\end{eqnarray*}
where, for $i=1,\cdots,n$, $\ell_i \triangleq \# \left\{ j: Y_{(j)} \geq Y_{(i)} \right\}$ and $r_i \triangleq \# \left\{ j: Y_{(j)} \leq Y_{(i)} \right\}$.

Thus, we define
\begin{eqnarray} \label{est-I}
\widehat{\mathcal{I}} = \left\{ k :   \widehat{\omega}_k \geq cn^{-\kappa} \ \text{for} \ k=1,\cdots,p  \right\}
\end{eqnarray}
as the estimated $\mathcal{I}$ that detects a set of important predictors, where $c$ and $\kappa \in (0,1/2)$ are prespecified threshold values. Such a screening procedure is called \textit{XI-SIS}. Moreover, as emphasized in Section~\ref{discuss-xi}, $\widehat{\omega}_k$ is able to measure the correlation for continuous or discrete response with the predictor $X_k$, it implies that (\ref{est-I}) is valid to identify important predictors for models (\ref{model}) and (\ref{logit}). This is one of main differences with the competitive methods (e.g., Li et al. 2012b; Fan and Lv 2008; Xia and Li 2020).

\subsection{Theoretical Results}

To see the validity of (\ref{est-I}), in this section we establish the theoretical property of the proposed
independence screening procedure built upon the new coefficient of correlation (\ref{xi-new-corr}).

\begin{theorem}
\label{theorem1}
Suppose that $Y$ is not a constant. There exist positive constants $c$, $K$, and $\kappa$ with $0 < \kappa < 1/2$, such that
\begin{eqnarray} \label{thm1-pr}
\text{pr}\left( \max \limits_{1\leq k \leq p} \left| \widehat{\omega}_k - \omega_k \right| > cn^{-\kappa} \right) \leq p \left[ \exp \left\{ -\log\left( 2cn^{1-\kappa}K/3 \right) \right\} + 3\exp\left(-2n^{1-2\kappa} c^2 K^2 /9 \right) \right].
\end{eqnarray}
Moreover, assume that $\min \limits_{1\leq k \leq p} \widehat{\omega}_k \geq cn^{-\kappa}$. Then we have the sure screening property
\begin{eqnarray*}
\text{pr}\left( \widehat{\mathcal{I}} \supseteq \mathcal{I} \right) \geq 1- q \left[ \exp \left\{ -\log\left( 2cn^{1-\kappa}K/3 \right) \right\} + 3\exp\left(-2n^{1-2\kappa} c^2 K^2 /9 \right) \right],
\end{eqnarray*}
where $q$ is the the cardinality of $\mathcal{I}$.
\end{theorem}

Similar to early work (e.g., Li et al. 2012b; Fan and Lv 2008; Sheng and Wang 2020; Mai and Zou 2013), Theorem~\ref{theorem1} implies that XI-SIS is able to select all the truly important predictors with an overwhelming probability. In addition, (\ref{thm1-pr}) indicates that the proposed method is able to handle the nonpolynomial
(NP) dimensionality with $p = o(\exp(n^{1-2\kappa}))$, which is similar to other model-free feature screening methods (e.g., Li et al. 2012b; Xia and Li 2020). On the other hand, different from other feature screening methods that may require additional assumptions to make the theory true, such as the uniformly subexponential tail probability (e.g., Li et al. 2012b) and boundness of density functions of $Y$ and $X_k$ (e.g., Wu and Yin 2015; Xia and Li 2020), the proposed method requires fewer conditions to derive the theoretical result. Finally, compare with Chatterjee (2020) who showed that $\widehat{\omega}_k$ converges almost surely to $\omega_k$ as $n \rightarrow \infty$, our result (\ref{thm1-pr}) gives the ``non-asymptotic'' result. In other words, with a large probability, $\left| \widehat{\omega}_k - \omega_k \right|$ is bounded above in terms of a finite sample size. Of course, when $n \rightarrow \infty$, (\ref{thm1-pr}) yields the consistency as shown by Chatterjee (2020).

%\begin{proof}[of Theorem~\ref{theorem1}]
%The proof should be here.
%\end{proof}

\section{Numerical Studies} \label{Num-study}

\subsection{Simulation Setup}

Let a $p$-dimensional predictor $X$ be generated from a multivariate normal distribution with mean zero and covariance matrix $\Sigma_X$ with entries being $0.5^{|i-j|}$ for $i,j=1,\cdots,p$. We consider $p=1000, 1500$ or $3000$. When $X$ is given, we consider the following nonlinear regression models:

\begin{description}
\item[M1:] $Y = 2X_1 + X_2^3 + 3\sin(8X_3) + \exp(X_4) + \epsilon$ with $\epsilon \sim N(0,1)$;

\item[M2:] $Y = 2\log X_1 + X_2^3 + \cos(8 X_3^2) + \sigma(X) \epsilon$ with $\epsilon \sim N(0,1)$ and $\sigma(X) = \sqrt{|X_1+X_2|}$ ;

\item[M3:] $Y = \left\{\left| X_1 + 0.5 \right| I(X_1 < 0) + \left| X_1 - 0.5 \right| I(X_1 \geq 0) \right\}  + 2X_2^3 + 3\cos(8 X_3^2) + \exp(-X_4) + \epsilon$, where $X_1$ is generated from a uniform distribution and $\epsilon$ follows a $t$-distribution with degree of freedom being one;

\item[M4:] $Y \sim \text{Bernoulli}(\pi(X))$ with $\pi(X) = \frac{\exp\left\{ X_1^3 + 3\sin(8X_2) + \exp(X_3) \right\}}{1+\exp\left\{ X_1^3 + 3\sin(8X_2) + \exp(X_3) \right\}}$.

\end{description}

Models M1-M3 produce the continuous response, and M4 yields the binary response. Regarding the first three models (M1-M3), M1 is a usual GAM model with the normal error term, M2 is a nonparametric regression model with heterogeneous errors (e.g., Wu and Yin 2015), and M3 is a GAM model with error term being other distribution. Most predictors in M1-M4 are nonlinear with the response. In addition, as commented in Chatterjee (2020), trigonometric functions (i.e., $\sin x$ and $\cos x$) and $\left\{\left| X_1 + 0.5 \right| I(X_1 < 0) + \left| X_1 - 0.5 \right| I(X_1 \geq 0) \right\}$ in M3 are oscillatory. 

We consider sample sizes $n=400$ and $600$, and thus, the artificial data are given by $\{(X_i,Y_i) : i=1,\cdots,n\}$. We repeat simulation 1000 times for each setting. As suggested in other references, we consider $d = \left[ \frac{n}{\log(n)} \right]$, where $[a]$ represents the integer part of $a$.

\subsection{Simulation Results}

Noting that our purpose is to identify important predictors, i.e., $X_1$--$X_4$ in M1 and M3, and $X_1$--$X_3$ in M2 and M4, from ultrahigh-dimensional data. To evaluate the finite sample performance of the proposed method, we follow the presentation similar to other relevant literature (e.g., Li et al. 2012b) to measure the frequency of picking up those important predictors. Specifically,  we compute the proportion that {\it each active predictor is selected}  out of 1000 simulations. Higher proportion indicates higher possibility that such a predicted could be detected. For the comparisons, we mainly examine the FanLv-SIS (Fan and Lv 2008) and the DC-SIS (Li et al. 2012b) methods because of the availability of R packages \texttt{SIS} and  \texttt{energy}, respectively. All numerical results are summarized in Tables~\ref{tab-M1}--\ref{tab-M4}.

In the presence of linear predictor, $X_1$ in M1 can be easily detected by three methods. However, when predictors have nonlinear relationships between the response, the FanLv-SIS method has the worst performance to identify important predictors. On the contrary, both DC-SIS and XI-SIS have higher proportions of selecting truly important predictors than the FanLv-SIS method. Compare with two different model-free methods, DC-SIS and XI-SIS, we observe that they have comparable performance if the predictor has a monotone trend, such as $\log (x)$ and $\exp (x)$, with respect to the response. However, when the trajectory is oscillatory, it is obvious to see that XI-SIS outperforms DC-SIS with higher proportion of identifying the truly important predictors. This result is consistent with the findings in Chatterjee (2020). Moreover, regarding the model M4 with the binary response, we can see that proportions of selecting the truly important predictors based on XI-SIS are higher than those based on DC-SIS. In general, from numerical results we conclude that XI-SIS can successfully detect the truly important predictors regardless of the trajectory of predictors as well as the distribution of responses.

\section{Data Analysis} \label{rda}

\subsection{Example 1: Analysis of Cardiomyopathy Microarray Data with Continuous Response}

In this section, we implement the proposed method to the cardiomyopathy microarray data, which were collected by Segal (2003). This dataset contains $p = 6319$ gene expressions and $n = 30$ specimens. The goal was to determine which genes were influential for overexpression of a G protein-coupled receptor, designated Ro1, in mice. The Ro1 expression level is denoted as the continuous response $Y$, and the predictors $X_k$’s are other gene expression levels.

\begin{figure}[!ht]
\centering
 \includegraphics[width=1.0\textwidth]{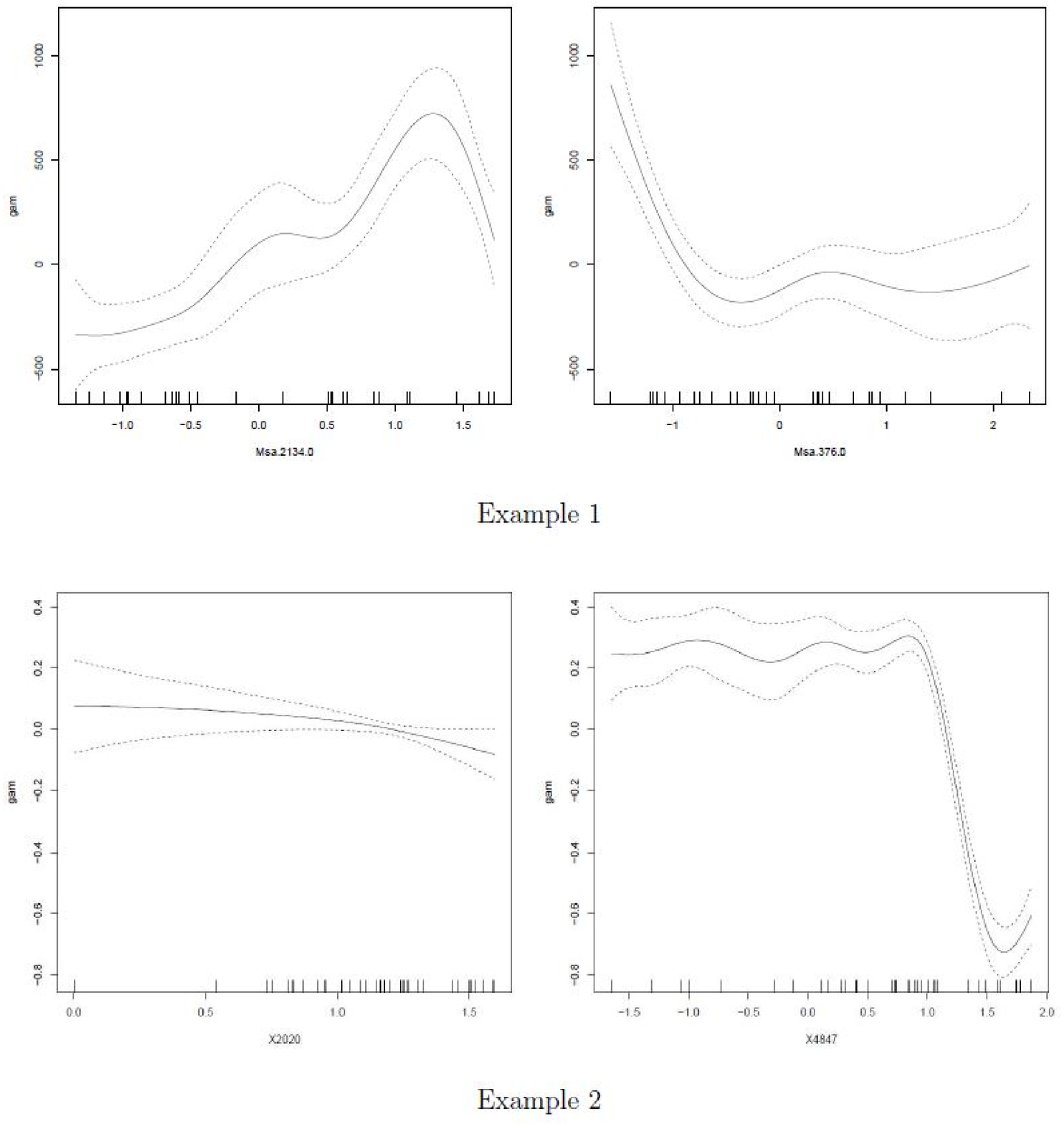}  
\caption{The curves obtained from GAM for two gene expressions detected by the proposed XI-SIS method. The solid curves are estimated curves and the dotted curves are 95\% confidence bands. Two above figures are based on Example 1, Two bottom figures are based on Example 2.}
\label{RDA-EX1}
\end{figure}

The two selected genes determined by the proposed XI-SIS are Msa.2134.0 and Msa.376.0. In contrast, DC-SIS selects Msa.2134.0 and
Msa.2877.0. Based on the dataset and two selected gene expressions, we fit the generalized additive model (GAM) with the continuous response (1), and display two estimated curves for two gene expressions in Figure~\ref{RDA-EX1}. We observe that both gene expressions have nonlinear relationship with the response. To see the fitness of GAM, we examine the adjusted $R^2$ and the deviance. With two selected gene expressions obtained by XI-SIS, the adjusted $R^2$ of the fitted model (1) is 97.7\% and the deviance
explained is 98.9\%, and the adjusted $R^2$ and the deviance for the DC-SIS method are 96.8\% and 98.3\%, respectively. We can see that the performance of the fitted models based on XI-SIS and DC-SIS are comparable. On the other hand, FanLv-SIS selects Msa.1166.0 and Msa.15405.0 that are totally different from the results of XI-SIS and DC-SIS.

Finally, to assess the prediction of fitted models, we adopt the $K$-fold cross-validation (e.g., Hastie et al. 2008, Section 7.10.1). Specifically, we split the data into $K=5$ roughly equal-sized parts. For each $k=1,\cdots,K$, let the $k$th part denote the testing data and let the remaining $K-1$ parts denote the training data. We first fit the training data with selected gene expressions, and then calculate the predicted values for the testing data. Let $\widehat{Y}_i^{(-i)}$ denote the predicted value for a subject $i$, computed with a subject $i$ in the $k$th part of the data removed. Then the cross-validate estimate of prediction error is defined as $\text{CV} = \sqrt{\frac{1}{n} \sum \limits_{i=1}^n \left( Y_i - \widehat{Y}_i^{(-i)} \right)^2}$. With gene expressions selected by XI-SIS, DC-SIS, and FanLv-SIS, the CV values are given by 280.374, 341.676, and 1245.180, respectively. It is clear to see that gene expressions selected by FanLv-SIS produce unsatisfactory prediction result. On the other hand, the prediction based on XI-SIS is slightly better than the result based on DC-SIS.

%\begin{figure}[!ht]
%\centering
%\begin{tabular}{c  c}
% \includegraphics[width=.4\textwidth]{RDA_EX1_2134.eps} & \includegraphics[width=.4\textwidth]{RDA_EX1_376.eps}
%\end{tabular}
%\caption{The scatter plot of $Y$ versus two gene expression levels identified by the proposed method.}
%\label{RDA-EX1}
%\end{figure}

%%%%%%%%%%%%%%%%%%%%%%%%%%%%%%%%%%

\subsection{Example 2: Analysis of Gene Expression Microarray Analysis with Binary Response}

Golub (1999) reported a gene expression microarray analysis expecting to identify
gene signature for the distinction between acute myeloid leukemia (AML) and acute
lymphoblastic leukemia (ALL), where gene expression levels were measured using
Affymetrix oligonucleotide arrays. The data contain 7128 genes and 72 specimens coming
from the two classes, with 47 specimens in class ALL and 25 specimens in class
AML. In particular, according to the study design, those 72 samples is composed of the
training data of 38 specimens (27 in class ALL and 11 in class AML) and the testing data of 34 specimens (20 in class ALL and 14 in class AML). In this study, the target response $Y$ is binary and the predictors $X_k$'s are gene expressions. The primary objectives are (a) to
identify the genes that are expressed differentially between AML and ALL, (b) to find possible pathways of genes that are expressed together, and (c) to classify the classes of AML and ALL using the selected gene expressions.

The two selected genes detected by the proposed XI-SIS are ID numbers 4847 and 2020, denoted as ``X4847'' and ``X2020''. On the other hand, it is interesting to see that DC-SIS and FanLv-SIS methods select the same gene expressions with ID numbers 2288 and 5772. Based on the training data and two selected gene expressions, we fit the GAM model with the binary response (2). Figure~\ref{RDA-EX1} displays the estimated curves for two gene expressions identified by XI-SIS. Different from ``X2020'' whose trend is relatively flattening, the trajectory of ``X4847'' is fluctuated, and thus, similar to the finding in simulation studies, both FanLv-SIS and DC-SIS do not detect ``X4847'' but the proposed XI-SIS does. In addition, with two gene expressions selected by XI-SIS, the adjusted $R^2$ of the fitted model (2) is 97.4\% and the deviance explained is 98.1\%, which are higher than the adjusted $R^2$ of 77.4\% and
the deviance explained of 80.4\% for the DC-SIS and FanLv-SIS methods. Since the adjusted $R^2$ values and the explained deviance are very large enough, it is no need to extract any additional gene expressions.

When the fitted model based on the training data is constructed, we further examine the prediction by using the testing data. Given the covariates $X_\text{new}$ in the testing data, let $\widehat{Y}_\text{new}$ denote the predicted classification based on the fitted model and let $Y_\text{new}$ be the true binary response in the testing data.

Define True Positive (TP), False Positive (FP), and False Negative (FN), respectively, as
\begin{eqnarray*}
\text{TP} &=& \sum \limits_{i=1}^{34} I(Y_{\text{new},i} = 1, \widehat{Y}_{\text{new},i} = 1), \ \
\text{FP} = \sum \limits_{i=1}^{34} I(Y_{\text{new},i} = 0, \widehat{Y}_{\text{new},i} = 1), \ \ \text{and} \\
\text{FN} &=& \sum \limits_{i=1}^{34} I(Y_{\text{new},i} = 1, \widehat{Y}_{\text{new},i} = 0).
\end{eqnarray*}
Ideally, a good prediction of classification needs large TP, and FP and FN should be as small as possible. To provide reasonably quantitative measures, we consider the F-measure to justify the performance of prediction.

We define {\it precision} (or called positive predictive value) and {\it recall} (or called true positive rate), respectively, as
\begin{eqnarray*}
\text{precision} =  \frac{\text{TP}}{ \text{TP}+\text{FP}} \ \ \text{and} \ \ \text{recall} = \frac{ \text{TP}}{\text{TP}+\text{FN}}.
\end{eqnarray*}
Then F-measure is defined as
\begin{eqnarray*}
\text{F-measure} = 2 \times \frac{\text{precision}\times \text{recall}}{\text{precision} + \text{recall}}.
\end{eqnarray*}
In principle, higher values of precision, recall and F-measure  reflect better performance of methods.

The prediction results of precision, recall, and F-measure are summarized in Table~\ref{tab-RDA-EX2}. We observe that those values based on XI-SIS are larger than those based on DC-SIS, showing that the fitted model based on XI-SIS yields more precise prediction than its competitive methods. It also verifies that two gene expressions ``X2020'' and ``X4847'' are important predictors in this dataset.

\section{Summary} \label{Summary}

In this paper, we extend the new formulation of the correlation coefficient proposed by Chatterjee (2020) to develop the feature screening procedure and detect the truly important predictors in ultrahigh-dimensional data. In addition to the equivalence of conventional Pearson's coefficient as mentioned in Chatterjee (2020), we also show that the point-biserial correlation is treated as a special case of (\ref{xi-new-corr}). Such a mathematical derivation justifies that (\ref{xi-new-corr}) is valid to measure two arbitrarily distributed random variables. With such a property, we develop the model-free feature screening method, which is easy to implement and is able to detect important predictors whose trajectories are nonlinear or oscillatory. We establish the sure screening property with required conditions fewer than other relevant literature to verify the validity of the proposed method.

Unlike some early methods that have been applied to other different settings, the correlation coefficient (\ref{xi-new-corr}) and the proposed method are newly developed approaches. As a result, it is interesting to extend the proposed method to complex settings, such as incomplete response induced by right-censoring, measurement error in predictors, or the iterated procedure to identify the falsely excluded predictors (e.g., Chen 2019). Those important topics are our future work.

\appendix

\begin{appendices}

\section{Proof of Theorem~1} \label{pf}

Let $\mathcal{F}(y) = \text{pr}(Y \geq y)$ and $F(y) = \text{pr}(Y \leq y)$ denote the ``survival function'' and ``cumulative distribution function (CDF)'' of $Y$, respectively. In addition, denote  $\widehat{\mathcal{F}} (y) = \sum \limits_{i=1}^n I(Y_i \geq y)$ and $\widehat{F} (y) = \sum \limits_{i=1}^n I(Y_i \leq y)$ as the estimators of $\mathcal{F}(y)$ and $F(y)$, respectively. Moreover, define
\begin{eqnarray} \label{est-omega-tilde}
\widetilde{\omega}_k 
= 1 - \frac{ \sum \limits_{i=1}^{n-1} \left| \widehat{F}(Y_{i+1}) - \widehat{F}(Y_i) \right|}{2 \sum \limits_{i=1}^n \widehat{\mathcal{F}}(Y_i) \left\{1 - \widehat{\mathcal{F}}(Y_i) \right\}  } 
\end{eqnarray}
and
\begin{eqnarray} \label{est-omega-ast}
\omega_k^\ast = 1 - \frac{ \sum \limits_{i=1}^{n-1} \left| F(Y_{i+1}) - F(Y_i) \right|}{2 \sum \limits_{i=1}^n \mathcal{F}(Y_i) \left\{1 - \mathcal{F}(Y_i) \right\}  }.
\end{eqnarray}

In the following derivation, we divide the proof into three steps.
\\
\textbf{Step 1:} Examine $\text{pr}\left( \left| \widehat{\omega}_k - \omega_k \right| > \delta \right)$ for some $\delta>0$. 

By the decomposition, we have
\begin{eqnarray} \label{decom-prob}
\text{pr}\left( \left| \widehat{\omega}_k - \omega_k \right| > \delta \right) 
&=& \text{pr}\left( \left| \widehat{\omega}_k - \widetilde{\omega}_k \right| > \delta/3 \right) + \text{pr}\left( \left| \widetilde{\omega}_k - \omega_k^\ast \right| > \delta/3 \right) + \text{pr}\left( \left| \omega_k^\ast - \omega_k \right| > \delta/3 \right) \nonumber \\
&\triangleq& A_1 + A_2 + A_3.
\end{eqnarray}
The remaining derivation is to examine $A_1$, $A_2$, and $A_3$ separately.
\\
\underline{Step 1.1:} We first examine $A_1$. 

According to the derivation of Chatterjee (2020), it can be shown that 
$-\sum \limits_{i=1}^{n-1} \left| \widehat{F}(Y_{i+1}) - \widehat{F}(Y_i) \right| = 2n \sum \limits_{i=1}^n \min\{ \widehat{F}(Y_i), \widehat{F}(Y_{N(i)}) \} - 2\sum \limits_{i=1}^n \widehat{\mathcal{F}}(Y_i) - \frac{r_n - r_1}{n}$, where $N(i) \triangleq \left\{\begin{array}{cc}
\pi^{-1}(\pi(i)+1) & \ \ \text{if }\ \  \pi(i) < n;  \\
i & \ \ \text{if } \ \ \pi(i) = n
\end{array}
\right.$ and $\pi(i)$ denotes the rank of $X_i$. As a result, (\ref{est-omega-tilde}) can be expressed as
\begin{eqnarray} \label{Qhat-Shat} 
\widetilde{\omega}_k 
= \frac{ n \sum \limits_{i=1}^n \min\{ \widehat{F}(Y_i), \widehat{F}(Y_{N(i)}) \} - \sum \limits_{i=1}^n \left\{\widehat{\mathcal{F}}(Y_i)\right\}^2 - \frac{r_n - r_1}{2n}}{\sum \limits_{i=1}^n \widehat{\mathcal{F}}(Y_i) \left\{1 - \widehat{\mathcal{F}}(Y_i) \right\}  } 
\triangleq \frac{ \widehat{Q}_n }{\widehat{S}_n}.
\end{eqnarray}
Based on the new representation (\ref{Qhat-Shat}), similar derivation in Chatterjee (2020) further yields
\begin{eqnarray*}
\left| \frac{ \widehat{Q}_n }{\widehat{S}_n} - \widehat{\omega}_k \right| \leq \frac{1}{2n \widehat{S}_n},
\end{eqnarray*}

Since the class of indicator functions is Glivenko-Cantelli, then by Glivenko-Cantelli theorem (e.g., van der Vaart and Wellner 1996), we have that 
$\widehat{\mathcal{F}}(y) \left\{1 - \widehat{\mathcal{F}}(y) \right\} \rightarrow \mathcal{F}(y) \left\{1 - \mathcal{F}(y) \right\} $  uniformly for all $y$ in the support of $Y$, denoted as $\mathcal{Y}$. In other words, $E\left( \widehat{S}_n \right)$ exists and is finite and nonzero. As a result, there exists some constant $K_0>0$ such that
\begin{eqnarray*}
E\left(\left| \widehat{\omega}_k - \widetilde{\omega}_k \right|\right) \leq E \left( \frac{1}{2n \widehat{S}_n} \right) \approx \frac{1}{2n E\left( \widehat{S}_n \right)} < \frac{1}{2n K_0}.
\end{eqnarray*}
Therefore, by the Markov inequality, we have
\begin{eqnarray} \label{ineq-A1}
\text{pr}\left( \left| \widehat{\omega}_k - \widetilde{\omega}_k \right| > \delta \right) &\leq& \frac{1}{\delta} \times \frac{1}{2nK_0} \nonumber \\
&=& \exp \left\{ -\log\left( 2nK_0\delta \right) \right\}.
\end{eqnarray}
\ \\
\underline{Step 1.2:} we examine $A_2$.

Continue the result and Glivenko-Cantelli theorem as mentioned in Step 1.1, we have $\frac{1}{\widehat{\mathcal{F}}(y) \left\{1 - \widehat{\mathcal{F}}(y) \right\}} \leq \frac{1}{\mathcal{F}(y) \left\{1 - \mathcal{F}(y) \right\}}$. Besides, by triangle inequality, we have
\begin{eqnarray*}
\left| \widehat{F}(Y_{i+1}) - \widehat{F}(Y_i) \right| - \left| F(Y_{i+1}) - F(Y_i) \right| \leq
\left| \widehat{F}(Y_{i+1}) - F(Y_{i+1}) \right| + \left| \widehat{F}(Y_i) - F(Y_i) \right|.
\end{eqnarray*}
Then we have
\begin{eqnarray} \label{diff-tilde-ast}
 \left| \widetilde{\omega}_k - \omega_k^\ast \right| 
 &=& \frac{1}{2} \left| \frac{ \sum \limits_{i=1}^{n-1} \left| \widehat{F}(Y_{i+1}) - \widehat{F}(Y_i) \right|}{\sum \limits_{i=1}^n \widehat{\mathcal{F}}(Y_i) \left\{1 - \widehat{\mathcal{F}}(Y_i) \right\}  } - \frac{ \sum \limits_{i=1}^{n-1} \left| F(Y_{i+1}) - F(Y_i) \right|}{\sum \limits_{i=1}^n \mathcal{F}(Y_i) \left\{1 - \mathcal{F}(Y_i) \right\}  }  \right| \nonumber \\
 &\leq & \frac{1}{2} \left| \frac{\sum \limits_{i=1}^{n-1} \left| \widehat{F}(Y_{i+1}) - \widehat{F}(Y_i) \right| - \sum \limits_{i=1}^{n-1} \left| F(Y_{i+1}) - F(Y_i) \right|}{\sum \limits_{i=1}^n \mathcal{F}(Y_i) \left\{1 - \mathcal{F}(Y_i) \right\}  }  \right| \nonumber \\
  &\leq & \frac{1}{2} \left| \frac{\frac{1}{n}\sum \limits_{i=1}^{n-1} \left| \widehat{F}(Y_{i+1}) - F(Y_{i+1}) \right| + \frac{1}{n}\sum \limits_{i=1}^{n-1} \left| \widehat{F}(Y_i) - F(Y_i) \right|}{\frac{1}{n} \sum \limits_{i=1}^n \mathcal{F}(Y_i) \left\{1 - \mathcal{F}(Y_i) \right\}  }  \right| \nonumber\\
    &\leq & \frac{1}{2K_1} \left| \frac{1}{n}\sum \limits_{i=1}^{n-1} \left| \widehat{F}(Y_{i+1}) - F(Y_{i+1}) \right| + \frac{1}{n}\sum \limits_{i=1}^{n-1} \left| \widehat{F}(Y_i) - F(Y_i) \right|  \right|,
\end{eqnarray}
where the last step comes from similar discussion in Step 1.1 and $K_1$ is a positive constant which satisfies $\frac{1}{n} \sum \limits_{i=1}^n \mathcal{F}(Y_i) \left\{1 - \mathcal{F}(Y_i) \right\} \geq K_1$. On the other hand, for given $\delta>0$, the Dvoretzky–Kiefer–Wolfowitz (DKW) inequality gives
$\text{pr} \left( \sup \limits_{y \in \mathcal{Y}} \left| \widehat{F}(y) - F(y) \right| > \delta \right) \leq 2 \exp\left(-2n \delta^2 \right)$. Therefore, together with (\ref{diff-tilde-ast}), we have
\begin{eqnarray} \label{ineq-A2}
 \text{pr}\left( \left| \widetilde{\omega}_k - \omega_k^\ast \right| > \delta \right) \leq 4 \exp\left(-2n K_1^2 \delta^2 \right).
\end{eqnarray}
\ \\
\underline{Step 1.3:} we examine $A_3$.

Similar to the discussion in Step 1.1, (\ref{est-omega-ast}) can be expressed as $\frac{ Q_n }{S_n}$, where $Q_n = n \sum \limits_{i=1}^n \min\{ F(Y_i), F(Y_{N(i)}) \} - \sum \limits_{i=1}^n \left\{\mathcal{F}(Y_i)\right\}^2 - \frac{r_n - r_1}{2n}$ and $S_n = \sum \limits_{i=1}^n \mathcal{F}(Y_i) \left\{1 - \mathcal{F}(Y_i) \right\}$.

By the strong law of large numbers, we have that as $n$ is large enough, $\frac{1}{n}S_n \rightarrow \int \mathcal{F}(t) \left\{1-\mathcal{F}(t) \right\}d\mu(t)$. In addition, since $ \left| \min\{ F(Y_i), F(Y_{N(i)}) \} - \left\{\mathcal{F}(Y_i)\right\}^2 - \frac{r_n - r_1}{2n}  \right| \leq 2$ and $Y_i$'s are assumed to be independent, applying Hoeffding's inequality gives that for any $\delta>0$,
\begin{eqnarray*}
 \text{pr} \left\{ \left| \frac{1}{n} Q_n - E\left( \frac{1}{n} Q_n \right) \right| \geq \delta \right\} \leq 2 \exp \left\{- \frac{2n^2\delta^2}{16} \right\},
\end{eqnarray*}
and thus, we conclude
\begin{eqnarray} \label{ineq-A3}
 \text{pr}\left( \left| \omega_k^\ast - \omega_k \right| > \delta \right) \leq 2 \exp \left\{- \frac{n^2 K_2^2 \delta^2}{8} \right\}
\end{eqnarray}
with a positive constant $K_2$ satisfying $\int \mathcal{F}(t) \left\{1-\mathcal{F}(t) \right\}d\mu(t) \geq K_2$

Finally, combining (\ref{decom-prob}) with (\ref{ineq-A1}), (\ref{ineq-A2}), and (\ref{ineq-A3}) yields
\begin{eqnarray} \label{target-pr}
\text{pr}\left( \left| \widehat{\omega}_k - \omega_k \right| > \delta \right) \leq \exp \left\{ -\log\left( 2nK_0\delta/3 \right) \right\} + \exp\left(-2n K_1^2 \delta^2/9 \right) + 2 \exp \left\{- \frac{n^2 K_2^2 \delta^2}{72} \right\}.
\end{eqnarray}
\ \\
\textbf{Step 2:} Examine $
\text{pr}\left( \max \limits_{1\leq k \leq p} \left| \widehat{\omega}_k - \omega_k \right| > cn^{-\kappa} \right)$ for some constants $c$ and $\kappa \in (0,1/2)$.

Take $K = \min\{K_0, K_1, K_2\}$, then (\ref{target-pr}) becomes
\begin{eqnarray*} 
\text{pr}\left( \left| \widehat{\omega}_k - \omega_k \right| > \delta \right) \leq \exp \left\{ -\log\left( 2nK\delta/3 \right) \right\} + 3 \exp \left\{- \frac{2 n K^2 \delta^2}{9} \right\}.
\end{eqnarray*}
Moreover, with $\delta$ replaced by $cn^{-\kappa}$, we have
\begin{eqnarray} \label{target-max-pr}
&&\text{pr} \left( \max \limits_{1\leq k \leq p} \left| \widehat{\omega}_k - \omega_k \right| > cn^{-\kappa} \right) \nonumber \\
%%%%%%%
&\leq & p \max \limits_{1\leq k \leq p} \text{pr} \left( \left| \widehat{\omega}_k - \omega_k \right| > cn^{-\kappa} \right) \nonumber \\
%%%%%%%
&\leq& p \left[ \exp \left\{ -\log\left( 2cn^{1-\kappa}K/3 \right) \right\} + 3\exp\left(-2n^{1-2\kappa} c^2 K^2 /9 \right) \right].
\end{eqnarray}
\ \\
\textbf{Step 3:} Prove sure screening property.

Recall that $\mathcal{I}$ and $\widehat{\mathcal{I}}$ are defined in Section~3.2. By (\ref{target-max-pr}), we can show that
\begin{eqnarray} 
 \text{pr} \left( \widehat{\mathcal{I}} \supseteq \mathcal{I} \right) 
&\geq&\text{pr} \left( \min \limits_{1\leq k \leq p} \left| \widehat{\omega}_k - \omega_k \right| > cn^{-\kappa} \right) \nonumber \\
%%%%%%%
&\geq & 1- q \text{pr} \left( \left| \widehat{\omega}_k - \omega_k \right| > cn^{-\kappa} \right) \nonumber \\
%%%%%%%
&\geq& 1- q \left[ \exp \left\{ -\log\left( 2cn^{1-\kappa}K/3 \right) \right\} + 3\exp\left(-2n^{1-2\kappa} c^2 K^2 /9 \right) \right],
\end{eqnarray}
where $q$ is the the cardinality of $\mathcal{I}$. Moreover, with $\kappa \in (0,1/2)$, when $n \rightarrow \infty$, we have $\text{pr} \left( \widehat{\mathcal{I}} \supseteq \mathcal{I} \right) \rightarrow 1$. It indicates that the estimated active set $\widehat{\mathcal{I}}$ includes the true active set that contains truly important predictors with probability approaching one. Therefore, the proof is completed. $\hfill \square$

\end{appendices}

\clearpage

 \begin{table}
       \huge
     \caption{Feature screening results for M1} \label{tab-M1}

   \scriptsize

 \centering
  \renewcommand{\arraystretch}{0.95}
 \begin{tabular}{cccccccccccccccc} 
 \\
 \hline
$n$  & $p$ &  \multicolumn{4}{c} { FanLv-SIS } & & \multicolumn{4}{c} { DC-SIS } & & \multicolumn{4}{c} { XI-SIS }\\ \cline{3-6} \cline{8-11} \cline{13-16} 

 &  &  $X_1$ & $X_2$ & $X_3$ & $X_4$  & & $X_1$ & $X_2$ & $X_3$ & $X_4$  & & $X_1$ & $X_2$ & $X_3$ & $X_4$
\\
 \hline 
400 & 1000 & 0.97 & 0.18 & 0.00 & 0.12 && 1.00 & 0.98 & 0.10 & 0.85 && 1.00 & 1.00 & 1.00 & 1.00\\
    & 1500 & 0.96 & 0.16 & 0.00 & 0.09 && 1.00 & 0.98 & 0.02 & 0.83 && 1.00 & 1.00 & 1.00 & 1.00\\
    & 3000 & 0.96 & 0.07 & 0.00 & 0.02 && 1.00 & 0.94 & 0.02 & 0.80 && 1.00 & 1.00 & 0.99 & 0.99 \\    \hline
600 & 1000 & 0.98 & 0.21 & 0.00 & 0.13 && 1.00 & 0.98 & 0.12 & 0.89 && 1.00 & 1.00 & 1.00 & 1.00\\
    & 1500 & 0.97 & 0.18 & 0.00 & 0.10 && 1.00 & 0.98 & 0.04 & 0.86 && 1.00 & 1.00 & 1.00 & 1.00\\    
    & 3000 & 0.96 & 0.10 & 0.00 & 0.04 && 1.00 & 0.95 & 0.03 & 0.84 && 1.00 & 1.00 & 1.00 & 1.00\\   
 \hline
\end{tabular}
\end{table}

 \begin{table}
       \huge
     \caption{Feature screening results for M2} \label{tab-M2}

   \scriptsize

 \centering
  \renewcommand{\arraystretch}{0.95}
 \begin{tabular}{cccccccccccccccc} 
 \\
 \hline
$n$  & $p$ &  \multicolumn{3}{c} { FanLv-SIS } & & \multicolumn{3}{c} { DC-SIS } & & \multicolumn{3}{c} { XI-SIS }\\ \cline{3-5} \cline{7-9} \cline{11-13} 

 &  &  $X_1$ & $X_2$ & $X_3$  & & $X_1$ & $X_2$ & $X_3$   & & $X_1$ & $X_2$ & $X_3$ 
\\
 \hline 
400 & 1000 & 0.01 & 0.01 & 0.00 && 0.96 & 0.98 & 0.01 && 0.98 & 0.98 & 0.99\\
    & 1500 & 0.00 & 0.00 & 0.00 && 0.96 & 0.97 & 0.00 && 0.97 & 0.97 & 0.98\\    
    & 3000 & 0.00 & 0.00 & 0.00 && 0.96 & 0.96 & 0.00 && 0.97 & 0.97 & 0.98 \\ \hline
600 & 1000 & 0.02 & 0.01 & 0.00 && 0.98 & 1.00 & 0.05 && 0.98 & 1.00 & 1.00\\
    & 1500 & 0.02 & 0.01 & 0.00 && 0.98 & 0.98 & 0.04 && 0.98 & 1.00 & 1.00\\    
    & 3000 & 0.01 & 0.01 & 0.00 && 0.97 & 0.97 & 0.04 && 0.98 & 0.98 & 0.99 \\ 
 \hline
\end{tabular}
\end{table}

 \begin{table}
       \huge
     \caption{Feature screening results for M3} \label{tab-M3}

   \scriptsize

 \centering
  \renewcommand{\arraystretch}{0.95}
 \begin{tabular}{cccccccccccccccc} 
 \\
 \hline
$n$  & $p$ &  \multicolumn{4}{c} { FanLv-SIS } & & \multicolumn{4}{c} { DC-SIS } & & \multicolumn{4}{c} { XI-SIS }\\ \cline{3-6} \cline{8-11} \cline{13-16} 

 &  &  $X_1$ & $X_2$ & $X_3$ & $X_4$  & & $X_1$ & $X_2$ & $X_3$ & $X_4$  & & $X_1$ & $X_2$ & $X_3$ & $X_4$
\\
 \hline 
400 & 1000 & 0.00 & 0.00 & 0.00 & 0.00 && 0.03 & 0.96 & 0.76 & 0.94 && 0.94 & 0.97 & 0.98 & 0.96\\
    & 1500 & 0.00 & 0.00 & 0.00 & 0.00 && 0.00 & 0.95 & 0.74 & 0.92 && 0.92 & 0.95 & 0.95 & 0.94\\    
    & 3000 & 0.00 & 0.00 & 0.00 & 0.00 && 0.00 & 0.91 & 0.70 & 0.90 && 0.90 & 0.94 & 0.94 & 0.94 \\    \hline
600 & 1000 & 0.00 & 0.02 & 0.00 & 0.01 && 0.04 & 0.96 & 0.80 & 0.94 && 0.95 & 0.97 & 0.97 & 0.96\\
    & 1500 & 0.00 & 0.01 & 0.00 & 0.00 && 0.01 & 0.96 & 0.78 & 0.94 && 0.95 & 0.96 & 0.96 & 0.96\\    
    & 3000 & 0.00 & 0.01 & 0.00 & 0.00 && 0.01 & 0.95 & 0.77 & 0.92 && 0.94 & 0.96 & 0.96 & 0.96\\    
 \hline
\end{tabular}
\end{table}

\begin{table}
       \huge
     \caption{Feature screening results for M4} \label{tab-M4}

   \scriptsize

 \centering
  \renewcommand{\arraystretch}{0.95}
 \begin{tabular}{cccccccccccccccc} 
 \\
 \hline
$n$  & $p$ &  \multicolumn{3}{c} { FanLv-SIS } & & \multicolumn{3}{c} { DC-SIS } & & \multicolumn{3}{c} { XI-SIS }\\ \cline{3-5} \cline{7-9} \cline{11-13} 

 &  &  $X_1$ & $X_2$ & $X_3$  & & $X_1$ & $X_2$ & $X_3$   & & $X_1$ & $X_2$ & $X_3$ 
\\
 \hline 
400 & 1000 & 0.08 & 0.00 & 0.03 && 0.90 & 0.10 & 0.93 && 0.97 & 0.97 & 0.96\\
    & 1500 & 0.05 & 0.00 & 0.02 && 0.90 & 0.07 & 0.92 && 0.96 & 0.97 & 0.96 \\
    & 3000 & 0.04 & 0.00 & 0.01 && 0.90 & 0.06 & 0.90 && 0.96 & 0.97 & 0.96 \\    \hline
600 & 1000 & 0.09 & 0.00 & 0.03 && 0.94 & 0.09 & 0.95 && 0.98 & 0.97 & 0.98 \\
    & 1500 & 0.03 & 0.00 & 0.03 && 0.93 & 0.07 & 0.94 && 0.97 & 0.96 & 0.97 \\    
    & 3000 & 0.02 & 0.00 & 0.02 && 0.93 & 0.04 & 0.94 && 0.97 & 0.96 & 0.97\\ 
 \hline
\end{tabular}
\end{table}

 \begin{table}
       \huge
     \caption{Prediction results for Example 2} \label{tab-RDA-EX2}

   \scriptsize

 \centering
  \renewcommand{\arraystretch}{0.95}
 \begin{tabular}{cccccccccccccccc} 
 \\
 \hline
& precision & recall & F-measure \\
DC-SIS & 0.850 & 0.944 & 0.895\\
XI-SIS & 0.950 & 0.950 & 0.950 
\\
 \hline

 \hline
\end{tabular}
\end{table}

\clearpage

\section*{References}

\refmark Chatterjee, S. (2020).
\newblock A new coefficient of correlation.
\newblock \textit{Journal of the American Statistical Association}, DOI: 10.1080/01621459.2020.1758115.

\refmark
Chen, L.-P. (2019).
\newblock Iterated feature screening based on distance correlation for ultrahigh-dimensional censored data with covariates measurement error.
\newblock \textit{arXiv:1901.01610v1}.

\refmark
Fan, J. and Lv, J. (2008). 
\newblock Sure independence screening for ultrahigh dimensional feature space (with discussion). 
\newblock {\em Journal of the Royal Statistical Society. Series B}, \textbf{70}, 849 -- 911.

\refmark
Fan, J., Samworth, R. and Wu, Y. (2009). 
\newblock Ultrahigh dimensional feature selection: beyond the linear model. 
\newblock {\em Journal of Machine Learning Research}, \textbf{10}, 1829 -- 1853.

\refmark
Fan, J. and Song, R. (2010). \newblock Sure independence screening in generalized linear models with NP-dimensionality. 
\newblock {\em The Annals of Statistics}, \textbf{38}, 3567 -- 3604.

\refmark
Fan, J., Feng, Y., and Song, R. (2011). \newblock Nonparametric independence screening in sparse
ultra-high-dimensional additive models.
\newblock {\em Journal of the American Statistical Association}, \textbf{106}, 544--557.

\refmark
Golub, L., Slonim, D., Tamayo, P., Huard, C., Gaasenbeek, M., Mesirov, J., Coller, H., Loh,
M., Downing, J., Caligiuri, M., Bloomfield, C., Lander, E. (1999).
\newblock Molecular classification of cancer:
class discovery and class prediction by gene expression monitoring.
\newblock \textit{Science}, {\bf 286}, 531--537.

\refmark
He, X., Wang, L., and Hong, H. G. (2013). \newblock Quantile-adaptive model-free variable screening for high-dimensional heterogeneous data. 
\newblock {\em The Annals of Statistics}, \textbf{41}, 342 -- 369.

\refmark
Li, G., Peng, H., Zhang, J., and  Zhu, L. (2012a).
\newblock Robust rank correlation based screening.
\newblock \textit{The Annals of Statistics}, 
\textbf{40}, 1846--1877.

\refmark
Li, R., Zhong, W., and  Zhu, L. (2012b).
\newblock Feature screening via distance correlation
learning.
\newblock \textit{Journal of the American Statistical Association}, \textbf{107}, 1129--1139.

\refmark
Mai, Q. and  Zou, H.
(2013). 
\newblock The Kolmogorov filter for variable screening in high-dimensional binary classification.
\newblock \textit{Biometrika}, \textbf{100}, 229--234.

\refmark
Segal, M. R., Dahlquist, K. D., and Conklin, B. R. (2003). 
\newblock Regression approach for microarray data analysis. 
\newblock {\em Journal of Computational Biology}, {\bf 10}, 961--980.

\refmark
Sheng, Y. and  Wang, Q.
(2020) 
\newblock Model-free feature screening for ultrahigh dimensional classification
\newblock \textit{Journal of Multivariate Analysis}, \textbf{178}, 104618.

\refmark
van der Vaart, A. W. and Wellner, J. A. (1996). 
\newblock  {\em Weak Convergence and Empirical Processes}. Springer, New York.

\refmark
Wood, S. N. (2017).
\newblock \textit{Generalized Additive Models: An Introduction with R}. CRC Press.

\refmark
Wu, Y. and Yin, G. (2015).
\newblock Conditional quantile screening in ultrahigh-dimensional heterogeneous data.
\newblock \textit{Biometrika}, \textbf{102}, 65--76.

\refmark
Xia, X. and Li, J. (2020).
\newblock Copula-based partial correlation screening: a joint and robust approach.
\newblock \textit{Statistica Sinica}. DOI: 10.5705/ss.202018.0219

\end{document}